\documentclass[12pt]{article}
\usepackage{graphicx} 
\usepackage{cite} 
\usepackage{multirow}%
\usepackage{amsmath,amssymb,amsfonts}%
\usepackage{amsthm}%
\usepackage{mathrsfs}%
\usepackage[title]{appendix}%
\usepackage{xcolor}%
\usepackage{textcomp}%
\usepackage{manyfoot}%
\usepackage{booktabs}%
\usepackage{algorithm}%
\usepackage{algorithmicx}%
\usepackage{algpseudocode}%
\usepackage{listings}%
\usepackage{geometry}
\newcommand{\bipin}[1]{\textcolor{blue}{#1}}

\begin{document}
	
	\begin{center}
		\textbf{\bipin{-- Preprint --}}
	\end{center}
	\begin{center}
		\textbf{Towards Location-Specific Precipitation Projections Using Deep Neural Networks}
	\end{center}
	
	\begin{center}
		Bipin Kumar$^1$, Bhvisy Kumar Yadav$^2$, Soumypdeep Mukhopadhyay$^3$, Rakshit Rohan$^3$, Bhupendra Bahadur Singh$^1$, Rajib Chattopadhyay$^{1,4}$, Nagraju Chilukoti$^2$, Atul Kumar Sahai$^1$
	\end{center}
	
	\begin{center}
		$^1$Indian Institute of Tropical Meteorology, Ministry of Earth Sciences, Government of India, Pashan, Pune 411008, India
	\end{center}
	
	\begin{center}
		$^2$ Earth and Atmospheric Sciences, National Institute of Technology Rourkela, Rourkela, Odisha, 769008, India
	\end{center}
	
	\begin{center}
		$^3$ Indian Institute of Science Education and Research, Pune, Dr. Homi Bhabh Road, Pashan Pune 411008, India
	\end{center}
	
	\begin{center}
		$^4$ India Meteorology Department, Pune, Shivaji Nagar, Pune 411005, India
	\end{center}
	
	\begin{center}
		Email: bipink@tropmet.res.in
	\end{center}
	
	\vspace{10pt}
	
	\begin{abstract}
		Accurate precipitation estimates at individual locations are crucial for weather forecasting and spatial analysis. This study presents a paradigm shift by leveraging Deep Neural Networks (DNNs) to surpass traditional methods like Kriging for station-specific precipitation approximation. We propose two innovative NN architectures: one utilizing precipitation, elevation, and location, and another incorporating additional meteorological parameters like humidity, temperature, and wind speed. Trained on a vast dataset (1980-2019), these models outperform Kriging across various evaluation metrics (correlation coefficient, root mean square error, bias, and skill score) on a five-year validation set. This compelling evidence demonstrates the transformative power of deep learning for spatial prediction, offering a robust and precise alternative for station-specific precipitation estimation.
	\end{abstract}

%
%
%
%
%



\section{Introduction}\label{sec:intro}

The ever-increasing deluge of data across diverse fields presents both opportunities and challenges in extracting meaningful information for practical applications. While vast datasets for example in climate science, foster profound insights, gaps in spatial coverage persist due to various constraints. These gaps impede comprehensive spatial analysis, and limit the applicability at local scale. Often sophisticated approximation techniques are used to bridge this gap. In meteorology, forecasts or projections are made by numerical models through solving relevant partial differential equations on discretized spatial grids. These models are globally used for operational forecast \cite{Rao2019,Willetts2017,Joseph2015}. Even the latest generation satellite precipitation, station based estimates, or merged records produce limited resolution data. Specifically, within a grid the patterns might remain overly uniform and lack spatial variations. Retrieving local scale information from such data leads to often erroneous estimates \cite{Huffman2001}. Typically, end-users require statistical information and forecasts at specific locations that often do not coincide with grid locations. Therefore, statistical methods are employed to approximate the values at these locations. Such estimations at a particular location are crucial from practical usage, for example, a flood forecaster may need rainfall quantity at some particular location or a city planner may need water level during heavy rainfall event for a certain locality which is more vulnerable for water logging. In such situations, quantitative information of meteorological variables is required to be estimated and disseminated at a specific location.

Traditionally, interpolation based  methods like Kriging \cite{Oliver1990} and other methods \cite{Osborn1997,Collins2013}, popular in geology and environmental sciences and related areas, dominated meteorological  applications for such geo-spatial location specific requirements \cite{Rajeevan2006,Cressie2015,Pai2014, Arsenault2014,Meng2021}. Kriging leverages weighted averages of data points, relying on the spatial covariance structure of the underlying process \cite{Matheron1963, Chen2022, Kumar2023}. However, limitations arise due to its assumption of stationary covariance, often violated in meteorological data, and computational bottlenecks for large datasets due to the need to invert an N$\times$N covariance matrix \cite{Chen2022}. This necessitates exploring alternative methods for more accurate, localized precipitation estimates \cite{Gaikwad2023}. One way is to perform the data downscaling using regional dynamical models which take the outputs from global models to produce dynamical downscaling \cite{Norbe2001} by taking the largescale atmospheric and oceanic conditions at the lateral boundaries \cite{Sanjay2020}. 
One notable drawback of this method is its dependency on the accuracy of global model fields and inherent biases, as highlighted by Seaby et al. \cite{Seaby2013}. Consequently, it necessitates bias correction before proceeding with additional downscaling steps. The second way is statistical data downscaling (SDD) \cite{Bano-Medina2020} which is based on machine learning techniques. 

Entering to the burgeoning field of Deep Learning (DL), it is understood that DL methods have revolutionized processes across various meteorological domains \cite{Reichstein2019,Rouet-Leduc2019,Ham2019}, including weather forecasting, statistical downscaling, and air pollution forecasting \cite{Dong2015, Vandal2017,Kumar2023, Singh2023,Kumar2022}. Shi et al. \cite{Shi2015} introduced an effective Convolutional Neural Network (CNN) based approach known as ConvLSTM for precipitation Nowcasting utilizing radar data. This method has subsequently been adopted by other researchers for forecasting various other variables \cite{Gaikwad2023}.
The CNN based and Generative Adversarial Network (GAN) based image high resolution methods have been used for the data downscaling for making high resolution data \cite{Kumar2022,Kumar2023}. There are several other methods and techniques used by the researches for statistical data downscaling of model and other data  \cite{Lafferty2021,Lafferty2023,Gutmann2022, Lange2019,Gergel2022, Thrasher2022,Kotamarthi2021,Wootten2020}.  The GAN methods, adept at learning to approximate a target distribution, have emerged as viable DL models that allow for the generation of realistic samples \cite{Goodfellow2014,Isola2017,Ravuri2021}. By learning stochastic functions, GANs can also capture small-scale variability that cannot be deterministically predicted, thus mitigating the issue of blurring often observed in neural network predictions \cite{Ravuri2021}.

\begin{figure*}[htpb]
	\centering
	\includegraphics[height=0.65\textheight, width=0.85\textwidth]{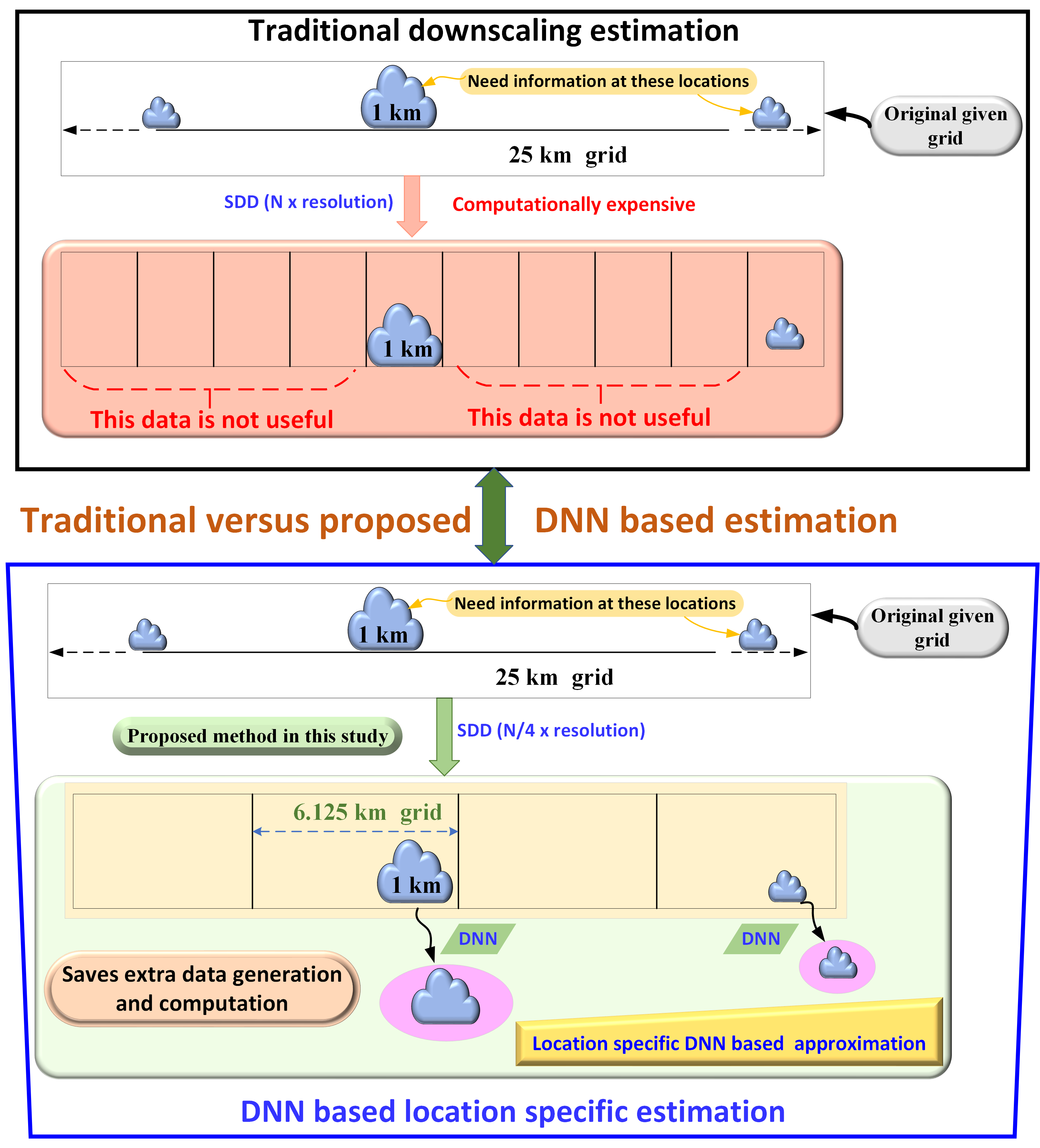}
	\caption{Depiction of the trade-off between computational expense and forecast precision, as well as minimizing superfluous data generation for providing location-specific forecast. }
	\label{fig:introduction}
\end{figure*}

In the context of precipitation, getting accurate location specific information e.g. of extremes based on a gridded data is difficult as the grids are usually coarser. Linear statistical method are utilized for operational purpose for example the Kriging method is used by the India Meteorological Department (IMD). 	A method based on Gaussian process regression, documented by Zhang et al. (2024) \cite{Zhang2021}, has been proposed as an alternative to traditional interpolation techniques to enhance the spatial resolution of sea surface temperature.   Although deep-learning (DL) techniques are employed for generating high resolution data (the statistical downscaling) as they progresses to higher resolutions, it becomes increasingly computationally demanding, prompting a need for data management costs.  This presents a trade-off between computational expense and forecasting accuracy. In meteorology, local-level variables are influenced by sub-grid scale phenomena like convective clouds of a few kilometers (say 5 km).  When considering a model grid of 100 kilometers, retrieving information at such small scales by applying downscaling methods described above to achieve resolutions up to sixteen times finer can become prohibitively expensive. Additionally, much of the data obtained at very high resolutions may not be practically useful as depicted in the figure \ref{fig:introduction}.

On the other hand, Deep Neural Networks (DNNs) can facilitate rapid analysis of massive datasets \cite{Najafabadi2015}. While CNNs excel at capturing spatiotemporal dependencies in image processing  \cite{Krizhevsky2012}, their extensive training needs and large feature space limitations often clash with the sparse, in-situ nature of many spatial approximation problems \cite{Chen2022}. 
 Spatiotemporal Graph Neural Networks (STGNNs) represent a category of methods with various versions that can be used for station data estimation. Jin et al. (2042) \cite{Jin2024} documented a comprehensive survey on these methods and their diverse applications across different fields.  They provide a survey on the variants of GNN methods, including physics-informed learning, and their applications in various filed including meteorology as well provided information about location based estimation of meteorological variables.

The DNN methods have been successfully applied for forecasting of flash flood or urban planning. In a study \cite{Ghimire2021} the application of fully connected DNNs is discussed  for the prediction of flush flood and for flood-prone area identification. It explores how DNNs handle multi-factor inputs like rainfall, topography, and soil saturation, which are critical for flash flood predictions. A comparison of machine learning method with DNN for flash flood is presented by Sit et al. \cite{Sit2019}. They documented that by integrating climate and soil data, the DNN model achieved high accuracy, outperforming traditional methods in predicting flash floods across different regions, especially in complex terrains. 

The DNN has been used for urban flash flood, where the model takes in inputs like rainfall, drainage, and land use data to predict flash floods in densely populated areas. The authors  demonstrate potential of DNNs in processing high-dimensional data for real-time flood prediction \cite{Feng2020}. A review by Masavi et al.  \cite{Mosavi2018} broadly covers machine learning techniques for flood prediction, it includes a section specifically on the applications of DNNs in flash flood forecasting. It details the advantages of DNNs in handling complex, multi-variable flood prediction problems and highlights specific case studies where DNN models significantly enhanced forecasting accuracy.

Long et al. (2024) \cite{Long2024} utilized a deep learning technique known as ConvLSTM to forecast drought propagation. Their model integrated multiple variables, including precipitation data obtained from the Standardized Precipitation-Evapotranspiration Index (SPEI), along with additional inputs from various sources such as Soil Moisture Index (SSMI). They offered insights across different regional scales, including forestland and cropland. However, the study did not provide information on location-specific predictions based on latitude and longitude. Studies have shown that DNNs significantly outperform traditional machine learning models in weather prediction tasks, including rainfall forecasting, due to their scalability and superior generalization capabilities \cite{Lecun2015, Rasp2018, Xie2020, Kumar2023} .

In this study, we propose and implement two novel NN models specifically designed for precise precipitation estimations at specific locations. These models also incorporate additional meteorological parameters to bolster the depth and accuracy of their estimation. Utilizing the curated dataset encompassing data from diverse sources, including the India Meteorological Department's (IMD) expansive network of over 6,000 rain gauges \cite{Rajeevan2006, Pai2014}, we train our models and validate them using various metrics. Our main goal is to systematically compare the performance of the established Kriging method and the DL-based models developed in this study. The significance of this research hinges on its potential to achieve unprecedented accuracy in localized precipitation estimates. 

\section{Data and Methodology}\label{sec:data}
The cornerstone of this research resided in the data, meticulously chosen to boost the learning and validation of our DNN models. IMD rain gauge values provided not only ground truth for comparison with Kriging but also crucial spatial context. Each data point has a daily temporal resolution, unveiling the intricate details of rainfall over time. The rainfall values, station elevation, latitude, and longitude are components of this data source. Furthermore, to expand our gaze beyond mere precipitation values, we tapped into the wealth of ERA5 ($0.25^o$) reanalysis data to enrich our models with additional meteorological parameters such as surface pressure, temperature, relative humidity, and wind profile which make a better representation of the atmospheric characteristics in the chosen dataset. The allocation of data was has been done with a good precision. Out of four decades (1980-2019) data, the 35 years data was served as the fertile ground for model training, providing ample nourishment for their learning algorithms. Table \ref{tab:data_details} documents the key characteristics of each data source, present its temporal and spatial resolutions.

Over the entire dataset, several notable meteorological events were observed, including 4 dry years, 9 years coinciding with El Niño, 4 years coinciding with La Niña, and 3 wet years \cite{Mol_iitm}. The dataset spanning five years, specifically 1995, 1996, 2001, 2013, and 2014, was selected for model testing. These five years constituted the crucible for validation, with each year handpicked to represent a distinct meteorological event. In particular, 2013 had abundant rainfall, while 2014 faced a deficit. 1995-1996, influenced by La Niña, challenged model adaptation, and 2001, a neutral year, served as a baseline for model evaluation.  Our main aim was to include monsoon variability in the unseen test dataset.

\begin{table*}
	\centering
	\begin{tabular}{|p{4.5 cm}|p{3.9 cm}|p{3.5cm}|}
		\hline
		\textbf{Data Variable}    & \textbf{Spatial Resolution}  & 	\textbf{Data Source} \\ 
		\hline
		Surface Pressure &		0.25 $\times$ 0.25 &	 ERA5    \\
		\hline
		Relative humidity &		0.25 $\times$ 0.25 &	 ERA5 \\
		\hline
		Temperature       &		0.25 $\times$ 0.25 &	 ERA5 \\
		\hline
		Zonal wind speed  &		0.25 $\times$ 0.25 &     ERA5   \\
		\hline
		Meridional wind speed &	0.25 $\times$ 0.25 &	 ERA5   \\
		\hline
		Station precipitation &	 At station        &	 IMD  \\
		\hline
		Grid precipitation  &	0.25 $\times$ 0.25 &	 IMD   \\
		\hline	
	\end{tabular}
	\caption{This table provides a comprehensive overview of the data utilized in this study. Each variable is represented at a daily resolution, constituting a high-frequency time series. The time duration for this data set spanning from 1980-2019.}
	\label{tab:data_details}
\end{table*}

\subsection{Methodology}\label{sec:methodlogy}
One of the main aims of this study is to find a better approach for approximating the data from grid points to specific locations over the Indian land region. We employed two main approaches namely, Kriging method (KM) and Neural Network (NN) model. The first one is generally used for the approximation of meteorological data from grid points to a specific locations \cite{Keegan-Treloar2021}. The KM stands as a powerful weapon in geostatistical methods, excelling in the art of spatial data prediction. Its strength lies in its ability to unveil the hidden tapestry of spatial dependence, weaving together observed data points with threads of statistical relationships.

The Kriging method is based on the interpolation technique hence has the possibility of numerical error. This numerical error can led do less correct or wrong estimations when applied to meteorological variables such as estimation for precipitation values from grid points to desired locations. Consequently, it can provide wrong forecast when the dynamical model forecasted output (gridded) is distributed at certain locations lie inside the grid. Therefore, a more accurate approximation approach is required to get more accurate estimation of meteorological variables and we considered the Neural Network approach for this purpose.. 

A Neural Network comprises an input layer, an output layer, and hidden layers, each housing N neurons tailored to project requirements. Given a set of N coordinates with available observations, the neural network seeks to determine the optimal predictor $\hat{y}^{opt} (S_0)$, predicting values at unobserved locations. The objective of spatial prediction is to identify the optimal predictor $\hat{y}^{opt} (S_0)$,  for the true process at an unobserved location $S_0$, based on observed values (Wanfang Chen, 2022). In decision theory, $\hat{y}^{opt} (S_0)$,   serves as the minimizer of an expected loss function or risk function (DeGroot, 2005).

\section{The NN Model and data pre-processing}\label{sec:model}

\begin{figure}[h]
	\centering
	\includegraphics[height=0.35\textheight, width=0.75\textwidth]{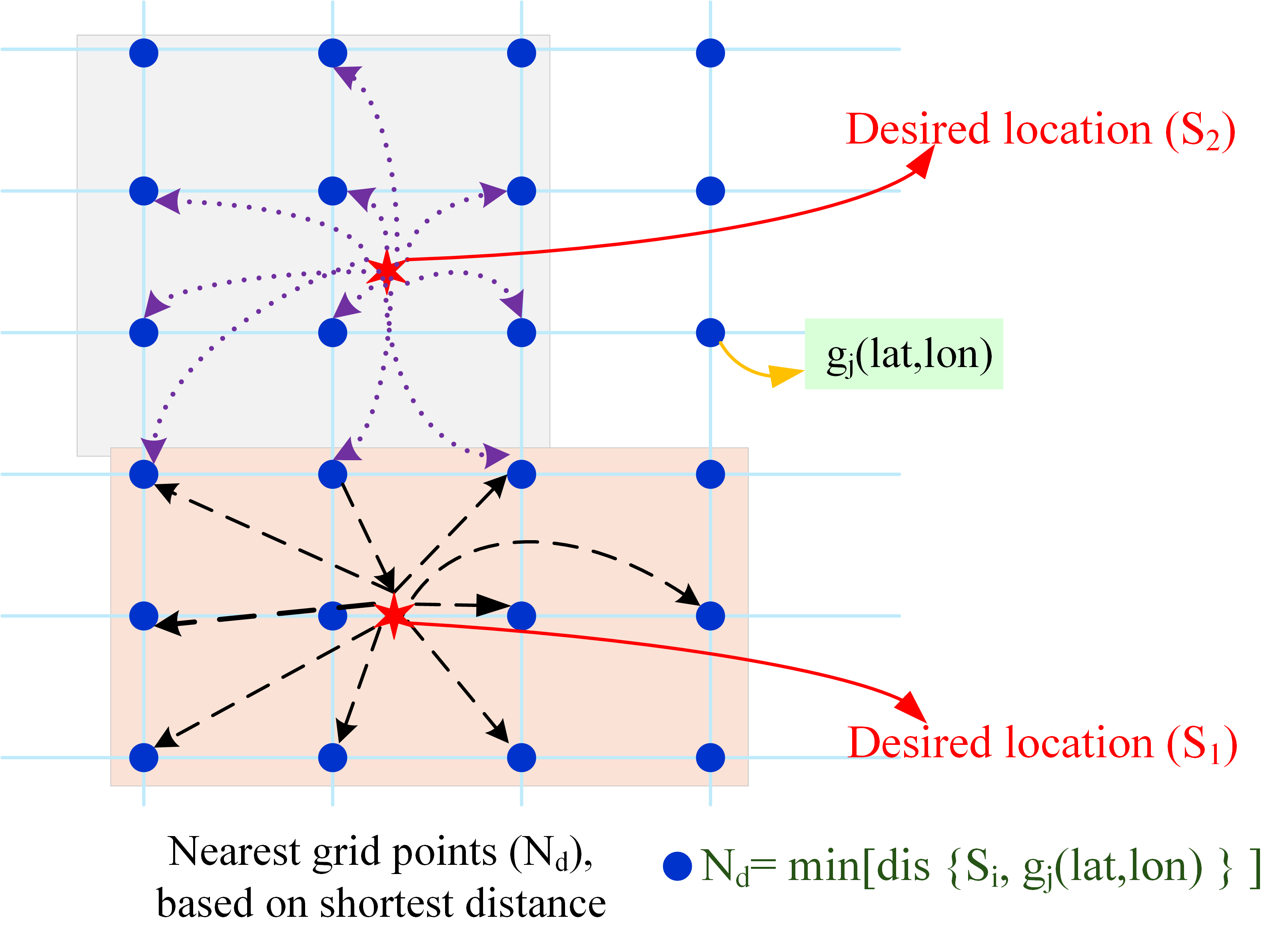}
	\caption{A representation of the nearest grid points surrounding the desired location. }
	\label{fig:neighbouring_grids}
\end{figure}

As stated before, for generating model training input data, we used two different sets of predictors. The first one is using only precipitation values at grid points and station  locations along with elevation and local coordinates, and in the second case, a few additional meteorological variables were considered. We constructed two types of input data files, each feeding information to a dedicated neural network model. The first, named `input file 1', focused on the immediate surrounding grid points of a chosen location as depicted in figure \ref{fig:neighbouring_grids}. We gathered data from the 10 nearest grid points, considering their latitude, longitude, elevation, and precipitation values. Additionally, we included the corresponding information for that location itself. This resulted in a total of 44 physical variables, 40 from the surrounding grid points and 4 from the location data. When training the model with this data, we used the 43 variables and location variables as input features, while the precipitation value at the station/location served as the ground truth label.

\begin{table*}
	\centering
	\begin{tabular}{|p{6.1 cm}|p{2.5 cm}|c|p{2.2 cm}|c| }
		\hline
		\textbf{Variables} &	\textbf{Input file1}& \textbf{ Status} &\textbf{Input file2} & \textbf{Status} \\
		\hline
		Precipitation at grid points       &  10 &  \checkmark  &  10 & \checkmark  \\
		\hline
		Precipitation at station location  &   1 &  \checkmark  &  1  & \checkmark  \\
		\hline
		Elevation                          & 11  &  \checkmark  & 11  &  \checkmark  \\
		\hline
		Latitude and Longitude        & 11$\times$2 =22 & \checkmark  &  11$\times$ 2=22 & \checkmark  \\
		\hline
		ERA5 meteorological  variables (listed in table 1)   & \hspace{3 mm} -- & -  & 10$\times$ 5=50 & \checkmark  \\
		\hline 
		Total variables      &   44 & & 94 & \\
		\hline
	\end{tabular}
	\caption{Details of the  data used for generating input files for the model training. }
	\label{tab:input_data_details}
\end{table*}

The second input file, `input file 2', expanded on the first by adding additional meteorological parameters alongside the existing four variables at each grid point. Five specific reanalysis variables (details are provided in table 1) were included in this dataset. This enriched the data considerably, bringing the total number of grid point variables to 9 and the overall variable count becomes (10*9+4)= 94, accounting for the station location data as well. The model trained on this data used the 93 grid point variables and location variables (excluding precipitation at location) as input features, with the  precipitation at that location once again serving as the label data. More illustrative details of the input file are provided in table  \ref{tab:input_data_details}.

To ensure optimal training of the model, we adopted early stopping criteria and finalized training at 200 epochs. To verify this choice, we conducted experiments with varying numbers of epochs, specifically testing at 150, 200, and 300 epochs. The selection of 200 epochs was based on achieving better convergence between the training and test loss. At this point, the test loss was minimal and closely aligned with the training loss, indicating that the model was neither underfitting nor overfitting. This decision aligns with the methodology recommended in recent research \cite{Nakazato2023, Nakazato2024}, where similar convergence-based criteria were suggested for determining the optimal number of epochs. 

\subsection{The Model Architectures }
We developed two separate neural network (NN) models, each tailored to its corresponding input data as discussed in previous subsection. Both models were trained on 80\% of the available data, reserving the remaining 20\% for validation purposes. The first model, handling the simpler "input file 1" data, had an input layer of 43 neurons. We employed PReLU activation functions in the hidden layers and leveraged the Adam optimizer to refine the model's parameters during training and the batch size was chosen 80.  The second NN model, processing the richer `input file 2' data, mirrored the first model's architecture in terms of hidden layers, activation functions, and optimizer. However, its input layer had 93 neurons, accommodating the increased number of features in the data. Again, the single output neuron predicted the precipitation value at desired location. To assess the performance of both models, we employed a set of relevant metrics, such as correlation coefficient (CC), root mean squared error (RMSE), bias calculated between ground truth and model output and skill score (S).

\section{Results and discussion}\label{sec:results}
This study aimed to find alternative approach to the interpolation for approximating a meteorological variable values at a specific location using the values at grid points. We employed DNN approach  for this purpose and compared the performance of traditional Kriging  Method (KM) and Neural Network (NN) models for estimating precipitation values at station (or desired) locations using grid point data. We have also compared the model's performance by a few ML models including XGBoost (results are not shown here). 
The results consistently demonstrated that NN models outperformed KM across all metrics. Notably, even the NN model without meteorological variables  achieved superior performance compared to KM, suggesting that the spatial relationships captured by the NN architecture itself were sufficient for accurate precipitation estimation. These findings highlight the potential of NN models as a powerful tool for precipitation interpolation, especially when limited meteorological data is available.

\subsection{Comprehensive analysis over the 5-years test data for monsoon season}

\begin{figure*}[htpb]
	\centering
	\includegraphics[height=0.72\textheight, width=0.92\textwidth]{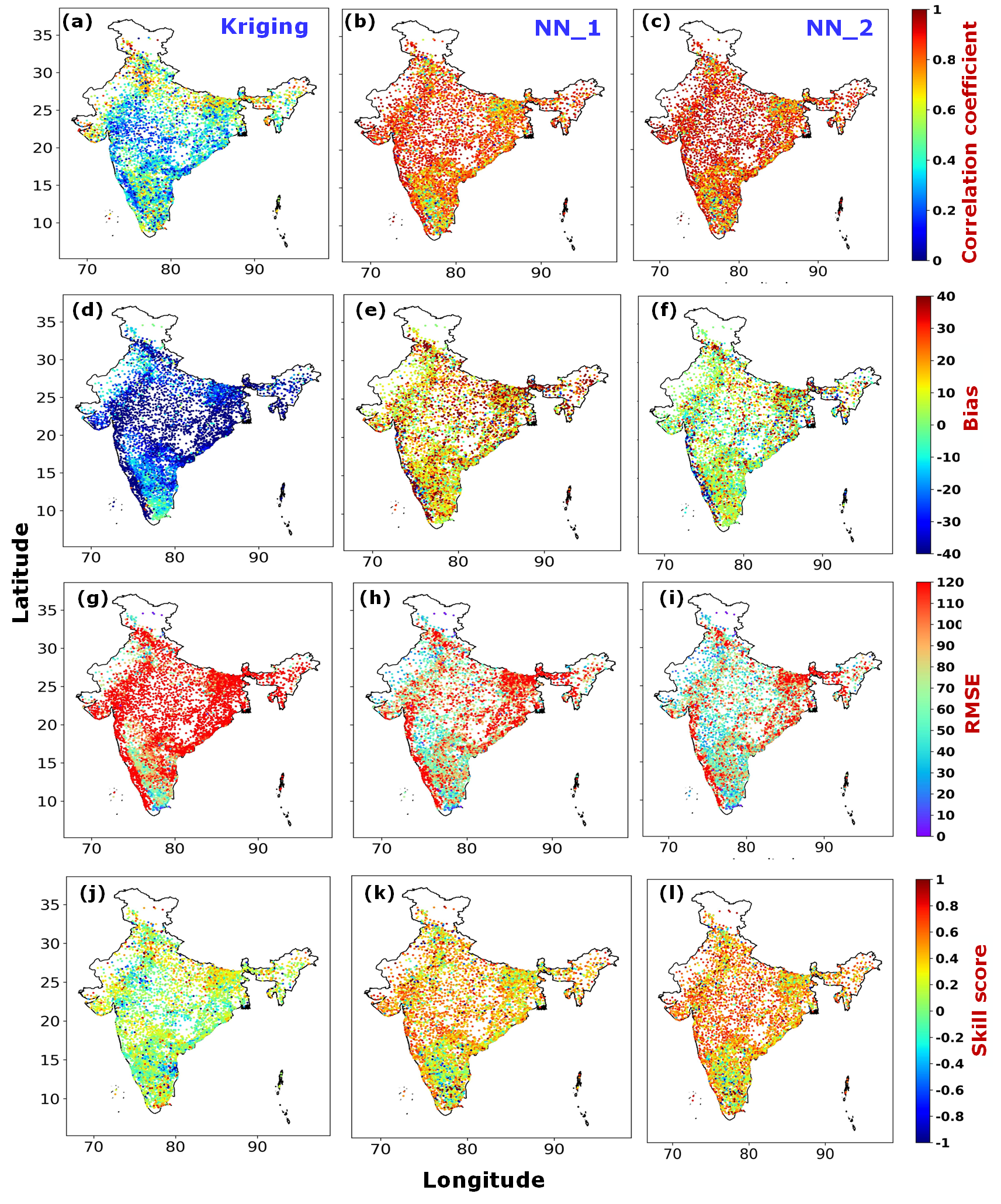}
	\caption{Representation of spatial plots CC, biases, RMSE and skill score obtained from three models. }
	\label{fig:spatial_plots_CC_RMSE_bias}
\end{figure*}

The performance of all models was comprehensively evaluated using a 5-year dataset spanning the Indian summer monsoon season (June-September, JJAS). Figure \ref{fig:spatial_plots_CC_RMSE_bias} presents the spatial distribution of key performance metrics for each model across the Indian landmass.
The KM (panel (a) in Figure \ref{fig:spatial_plots_CC_RMSE_bias}) exhibits significantly lower correlation coefficient (CC) compared to both NN models (panels (b) and (c)). This observation suggests that Kriging struggles to capture the spatial variations in precipitation observed across most stations. The NN\_Model\_2 consistently outperforms NN\_Model\_1 in terms of CC, indicating its superior ability to capture the spatial intricacies of rainfall. Similar trends are observed in the bias analysis (panels (d)-(f)). Kriging exhibits negative and high biases in coastal areas and central India compared to the NN models. The NN\_Model\_2 displays the least bias across most stations.
Panels (g)-(i) visualize each model's spatial distribution of the RMSE. Following the CC analysis, Kriging demonstrates higher RMSE values than both NN models, reflecting lower accuracy. Notably, all models exhibit relatively high RMSE at specific locations.  To further assess the model performance, the skill score (SS) \cite{Wilks2006}  was calculated which is defined  as 	SS = 1 - $\frac{RMSE}{\sigma}$. 

In this equation , the $\sigma$ stands for standard deviation of the data used for model training. 
Higher positive SS values indicate better model performance. Kriging showed lower values in most locations, ranging from -1.0 to 0.4, with some higher values at specific points. Conversely, the NN models displayed positive and higher SS values, except for a few locations with negative scores.
NN\_Model\_2 outperforms both Kriging and NN\_Model\_1 in all four metrics (CC, bias, RMSE, and SS), consistently exhibiting the lowest error across most stations. This analysis reinforces the superior performance of NN\_Model\_2 in capturing the spatial variability of precipitation. However, Figure \ref{fig:spatial_plots_CC_RMSE_bias} reveals interesting geographical variations in the models' RMSE values. Further investigation is needed to understand the factors contributing to higher RMSE in the specific regions linked to the complex terrain and orographic effects of the Western Ghats, diverse climates and limited data in Central India, and the mountainous terrain with complex weather dynamics in Northeast India all contribute to higher RMSE values across models.

\begin{figure*}[htpb]
	\centering
	\includegraphics[height=0.40\textheight, width=0.90\textwidth]{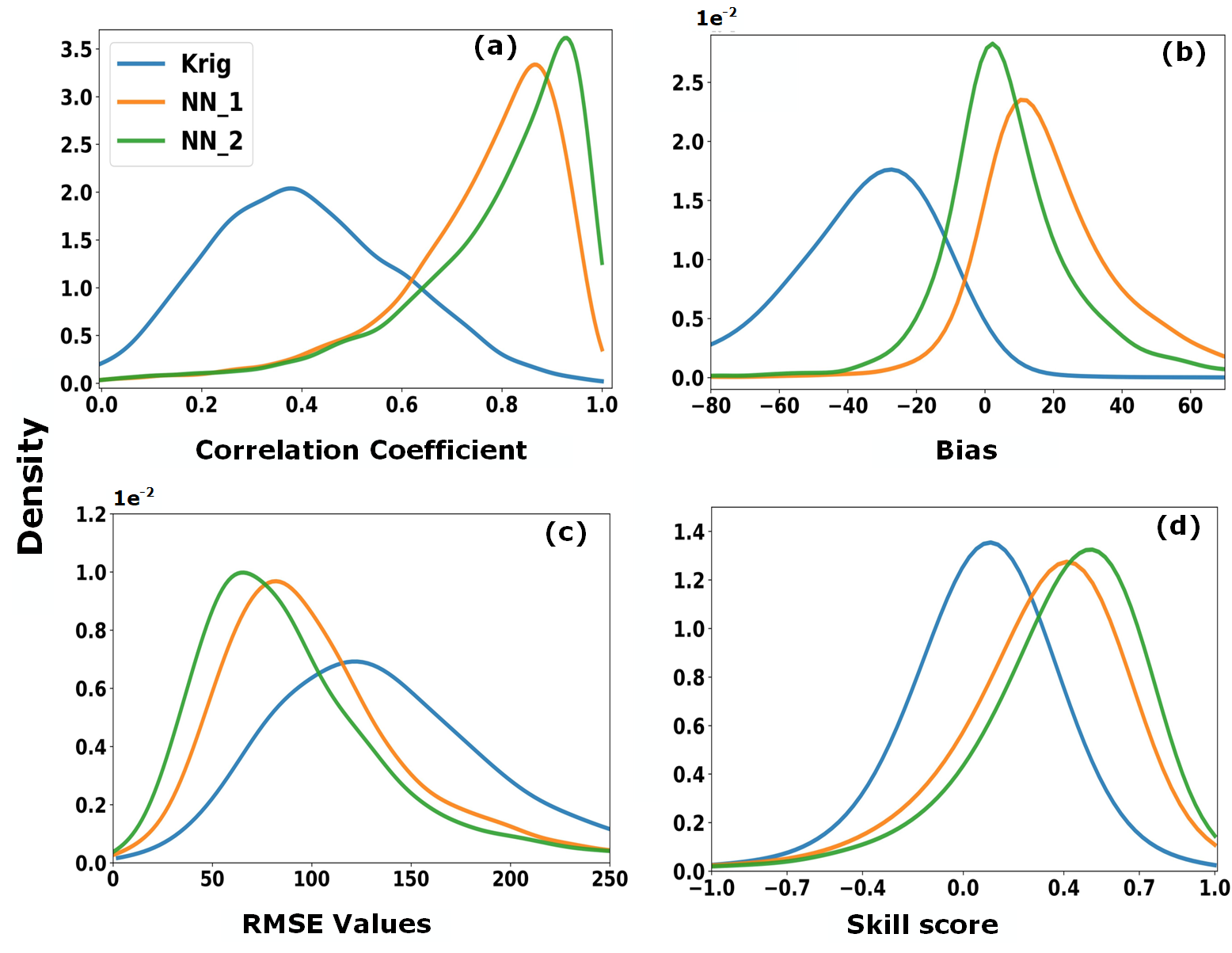}
	\caption{Performance evaluation of three methods using density plots for 5-years test data. }
	\label{fig:test_data_density_plots}
\end{figure*}

The skill score analysis (SS) enables a comparative assessment of the variability across different locations. This is evident in Figure~\ref{fig:spatial_plots_CC_RMSE_bias}, panels (j)-(l), which demonstrate that SS values exhibit relatively consistent levels across most regions. On the contrary, the RMSE displays higher values in specific locations. These  performance of the SS metric suggests that regions with higher RMSE values are characterized by greater variability within the precipitation data.
Building upon the spatial analysis that identified NN\_Model\_2 as the leading contender, we delved deeper by examining probability density functions (PDFs) for four key metrics: correlation coefficient (CC), root mean squared error (RMSE), bias between ground truth and model values, and skill score (SS). These PDFs, presented in Figure \ref{fig:test_data_density_plots}, illuminate the strengths and nuances of each model's performance.

Panel \ref{fig:test_data_density_plots}(a) illustrates CC values for NN\_Model which is around 0.8-0.9 across most stations, indicating a remarkably strong correlation with ground truth data. The density of biases depicted in panel \ref{fig:test_data_density_plots}(b) shows that Kriging method mostly produces the negative biases while the NN models have positive biases.  Similar dominance of the NN models  is observed in panel (c) for RMSE, where NN\_Model\_2 achieves the lowest values, implying the closest fit to ground truth on average. Panels  3(d) depict a similar picture for SS score further solidifying NN\_Model\_2's superior ability to capture the intricacies of precipitation patterns. Here, the Kriging has negative skill scores for many places and has the high peak at 0.0 score. 
While NN\_Model\_2's overall performance shines through, it's crucial to acknowledge the potential for contextual variations in accuracy. This is evident in the observed fluctuations for all models, including the two NNs and Kriging, underscoring the importance of considering specific geographical challenges when choosing the optimal method. In general, NN\_Model\_2 emerges as a powerful tool for generalizing across diverse locations, but its effectiveness can be further enhanced by tailoring strategies to address the unique complexities of specific regions.

We first conducted an analysis specifically for the monsoon season, considering that 70-80\% of the rainfall in the Indian subcontinent occurs during this period. To gain further insights into the model's performance, we extended our investigation to encompass the entire year, evaluating data for all 12 months. The same metrics used during the monsoon season analysis were calculated for each test year independently. 	
In the supplementary section, figure \ref{fig:spatial_plots_CC_RMSE_bias_whole_year} illustrates the values of these metrics. Remarkably, the overall performance of the models remains consistent with that observed during the monsoon season. Specifically, NN\_Model\_2 continues to outperform the other models, exhibiting higher correlation coefficients (CC) and lower RMSE and biases. 	
However, it's important to note that the quantitative values differ, indicating a relatively lower performance for all three models across these metrics. Further, the Kriging method demonstrates higher correlation, lower RMSE, and reduced negative biases compared to its performance during the monsoon season. A similar trend is observed for both NN models. This discrepancy may be attributed to the limited quantity of rainfall data available throughout the entire year.

\subsection{Better performing locations}
\begin{figure*}[htpb]
	\centering
	\includegraphics[height=0.22\textheight, width=0.99\textwidth]{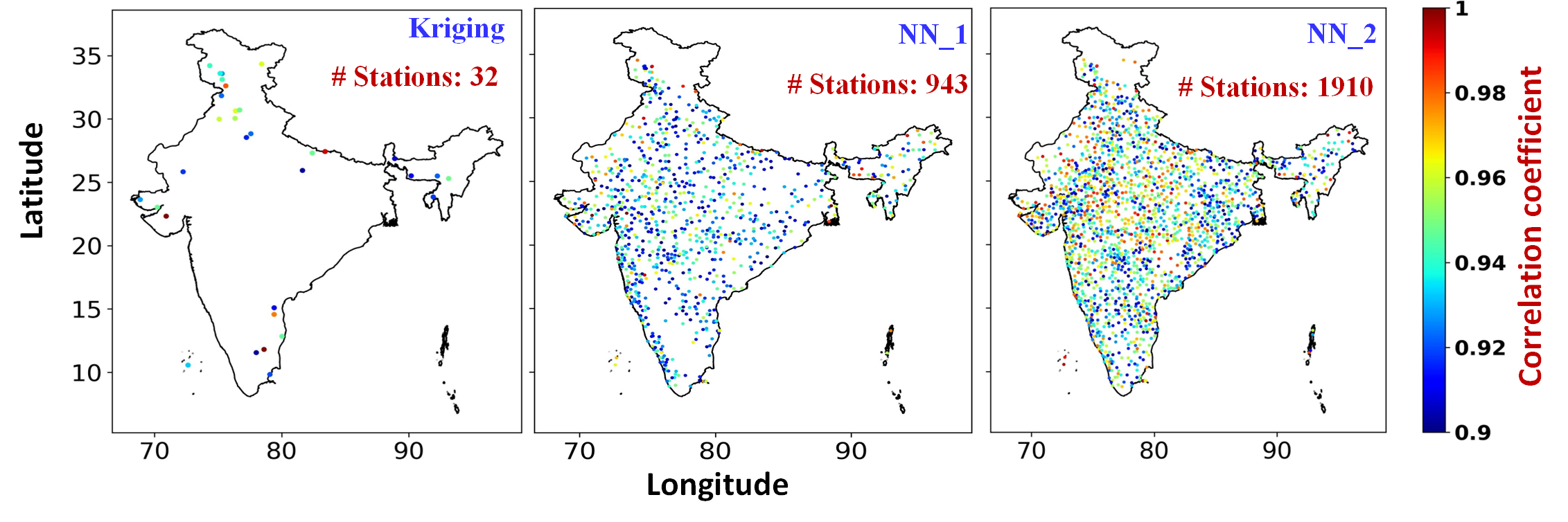}
	\caption{The station locations where the model performed exceptionally well, yielding very high correlation values. The total count of stations where the correlation coefficient (CC) value exceeded the threshold of 0.90 is provided for all three models are also indicated. }
	\label{fig:high_CC}
\end{figure*}

The plots presented in Figure \ref{fig:spatial_plots_CC_RMSE_bias} and Figure \ref{fig:test_data_density_plots} offer comprehensive insights into the performance of the models across the entire Indian landmass. Observations from the scatter plots reveals discernible patterns of model efficacy at specific geographical locations, thereby necessitating a detailed investigation to ascertain the nuances of individual model performance. Subsequently, we imposed a threshold CC value of 0.90 and conducted an exhaustive analysis to identify all stations where the models exhibited CC values surpassing this threshold. The geographic distribution of these stations, along with the aggregate count, is illustrated in Figure \ref{fig:high_CC}. This investigation underscores the robustness of NN\_Model\_2, which demonstrated superior performance by making the highest count of stations crossing the assigned CC threshold value.
The similar analysis of high CC value has been conducted for using whole year data which is depicted in the figure \ref{fig:high_CC_whole_year}  illustrate the stations location and the total station count where the CC  cross the threshold vale by all the models.

\subsection{Year-wise data analysis for whole year data}
While analyzing the entire 5-year dataset provided valuable insights into our models' overall performance, we sought to further understand their strengths and weaknesses under specific meteorological conditions. To achieve this, we carried out a detailed analysis of their performance on each year using the whole year data within the dataset: 1995, 1996, 2001, 2013, and 2014. Each of these years holds a unique meteorological significance, as outlined in Section \ref{sec:data}, serving as a crucial test bed for our models' adaptability and robustness.
\begin{figure*}[htpb]
	\centering
	\includegraphics[height=0.75\textheight, width=0.90\textwidth]{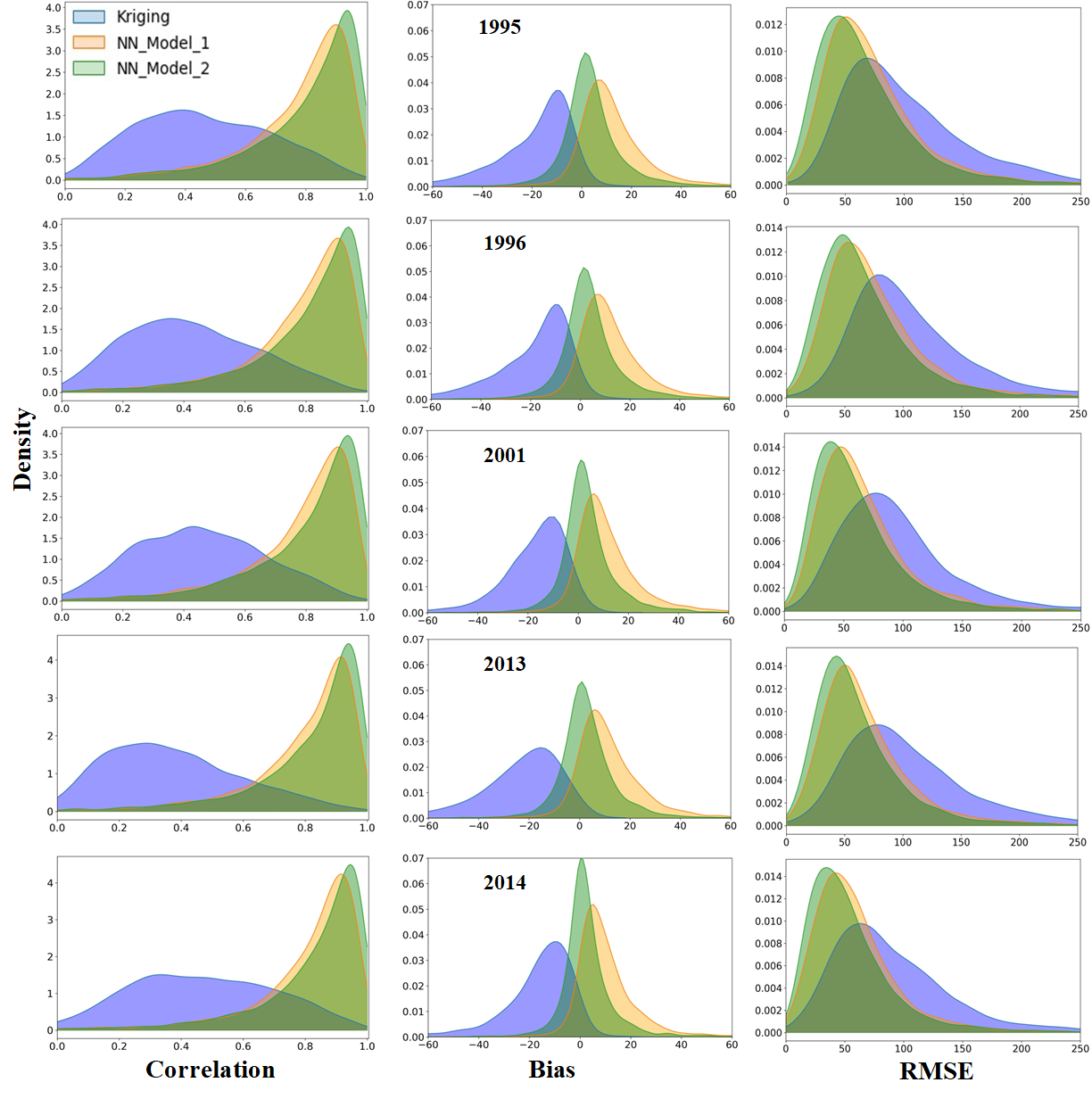}
	\caption{Density plots utilizing complete year data to demonstrate the comparative performance of all models based on correlation coefficient (CC), bias, and root mean square error (RMSE).  }
	\label{fig:yearwaise_data_density_plots}
\end{figure*}
Figure \ref{fig:yearwaise_data_density_plots}  provides a vivid picture of the year-wise performance through density plots for CC, biases and RMSE. It reveals that the NN models emerged as champions of consistency, and show remarkable stability across all years. Their CC values remained lies in the impressive range of 0.8 to 1.0, regardless of the varying meteorological conditions each year presented. This unwavering performance serves as a testament to their ability to generalize effectively, adapting to diverse weather patterns without sacrificing accuracy.  However, the Kriging method presented a contrasting narrative. Its performance, while capable at times, varies with the specific year's meteorological symphony. The year 2013, resulted in the lowest CC value amongst all years. Conversely, 2001, with its more predictable weather patterns, allowed Kriging to shine,
producing its highest CC value. This variability, in stark contrast to the NN models' unwavering resolve, highlights the potential limitations of Kriging in situations with significant meteorological shifts.  It is notable that the Kriging has produced the negative biases in all the years while this is not case for both NN models. 

The similar situation continues with RMSE, mirroring the contrasting trajectories of the models. While the NN models maintained their stoic position, delivering consistently low RMSE values across all years, Kriging once again revealed its susceptibility to the year's meteorological whims. The year 1995, with the influence of La Nina, allowed Kriging to achieve its lowest RMSE, while 1996 which also had the same influence, resulted in its highest. This year-wise analysis unveils the NN models' hidden strength: their remarkable resilience in the face of diverse meteorological landscapes. Therefore, this deeper dive into individual years solidifies our confidence in the NN models as the champions of consistency. Their unwavering performance, regardless of the year's meteorological peculiarities, positions them as reliable tools for accurately estimation of meteorological variables at a specific locations.

 Another advantage of using DNN is that  it outperforms in computational efficiency and scalability when dealing with large datasets, making it more suitable for high-dimensional meteorological data. Once trained, DNN models can quickly estimate spatial fields at any desired location, as the forward pass is computationally inexpensive. On the other hand, Kriging requires recalculating weights and solving large linear systems with new data points, which limits scalability and significantly slows performance as datasets grow.  The limitation of a DNN model lies in its strong dependency on the availability and quality of data; having more data generally improves the model's performance. However, data sparsity presents a  challenge, which is not unique to DNNs but is also a common issue in methods like Kriging and other deep learning techniques. The uncertainties in DNN predictions arise from two key sources: epistemic uncertainty, related to the model's limitations and lack of knowledge, and aleatoric uncertainty, stemming from inherent noise and variability in the data. In future, one can adopt the latest technologies, such as differentiable programming to exploit sensor placement within the training of a neural network model and the Senseiver approach for reconstructing complex spatial fields from few observations with low overhead \cite{Santos2023, Marcato2024} and apply them to different datasets. 

\section{Conclusion}\label{sec:conclusions}
This study explored the crucial issue of accurately approximating meteorological variables like precipitation values from grid points to specific locations with the main goal of enhancing the dissemination of weather/climate forecast model outputs to hyper local estimations. To achieve this, we introduced a novel approach by developing Deep Neural Network (DNN) models as alternatives to traditional interpolation methods such ordinary Kriging. We conducted a comprehensive performance comparison with the Kriging method. The entire research workflow, from data collection to model performance evaluation, is visually summarized in Figure \ref{fig:flowchart}.

Two distinct NN models were formulated: NN\_Model\_1, incorporating precipitation values, elevation, and location coordinates, and NN\_Model\_2, which expanded on the same features with the inclusion of five additional meteorological variables, namely, surface pressure, relative humidity, temperature, zonal wind speed and meridional wind speed.
Incorporating these additional meteorological variables improves accuracy due to the dependency of the precipitation of several factors. Topography affects precipitation through orographic lifting, humidity influences cloud formation and rainfall intensity, temperature impacts evaporation and cloud formation, and wind speed determines moisture movement and precipitation patterns. Including these variables enables more precise, location-specific precipitation estimations and provide better performance of the model in terms of metric values. In particular, the max correlation coefficient (CC) value in the pdf density plot (Fig 4)  moved from 0.81 to 0.9, the max value of RMSE in the same figure reduced from 90 to 53 for the second model. Similarly, the skill score peak increases from 0.4 in the first NN model to 0.6 in the second model. 

 The models were developed using a 40-year dataset (1980-2019), with 35 years designated for training and specific test years (1995, 1996, 2001, 2013 \& 2014) selected for their meteorological significance. Comprehensive and year-wise analyses were conducted to assess model performance under varying conditions. The model performance evaluation, utilizing metrics such as pattern correlation, RMSE, bias, and skill score, consistently demonstrated the superiority of the DNN models, even with the simpler feature set of NN\_Model\_1. These findings suggest that DNNs offer a promising solution to overcome the limitations of traditional interpolation techniques. The study provided two key insights:

\begin{figure}[htpb]
	\centering
	\includegraphics[height=0.68\textheight, width=0.85\textwidth]{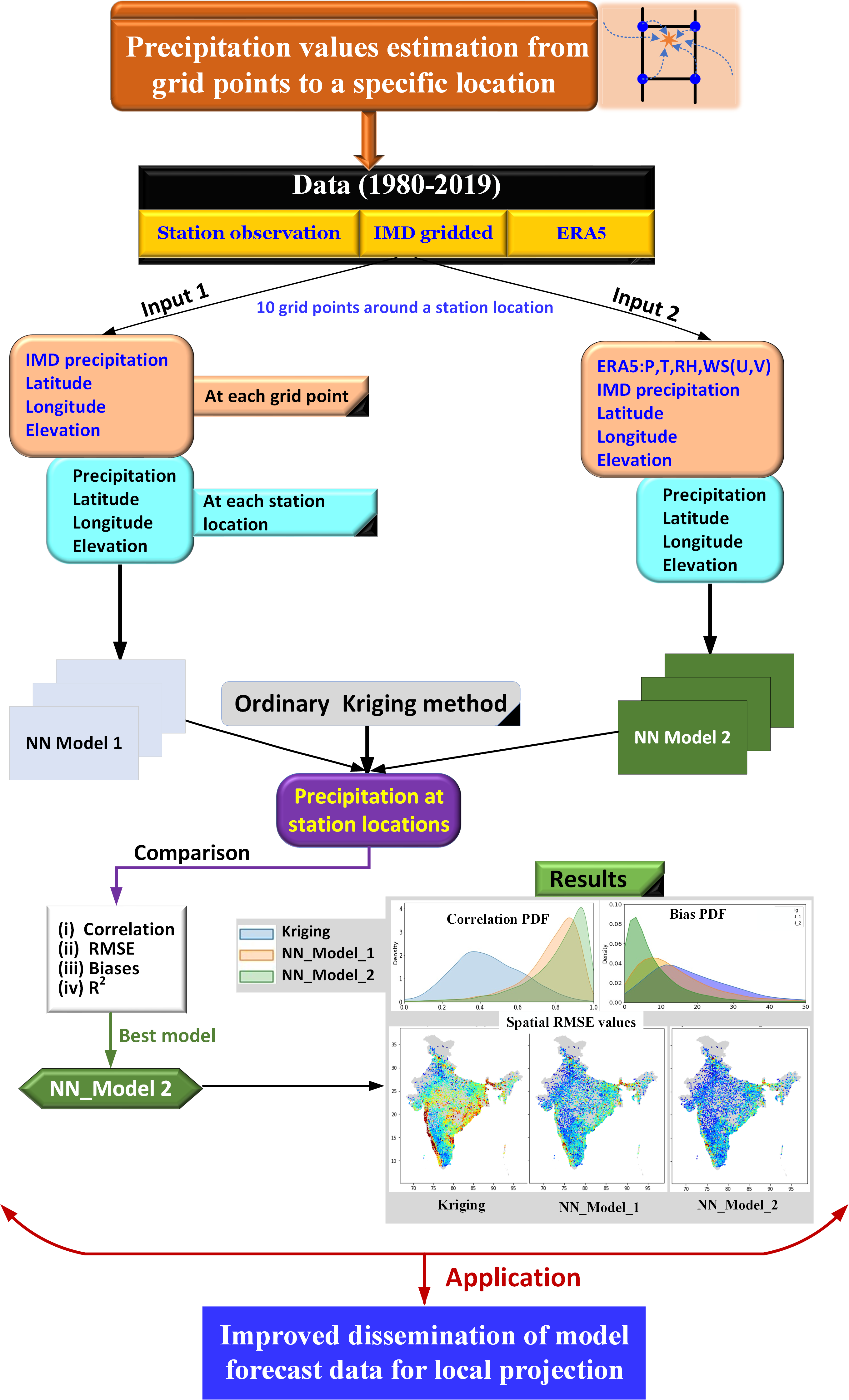}
	\caption{Conceptual diagram illustrating the work done in this study.  }
	\label{fig:flowchart}
\end{figure}

\begin{enumerate}
	\item[(i)]	A thorough examination of DL-based model performance at both station-level and the whole-dataset addressed geographical variations in accuracy, providing a robust understanding of the model`s strengths and weaknesses. This analysis aids in selecting the most suitable tool for a specific region. 
	\item[(ii)]	The study enhances spatial analysis of meteorological variables and highlights deep learning's potential in meteorology, offering a valuable alternative for improved spatial approximation.
\end{enumerate}

Future research could explore the scalability of this approach to different regions and its adaptability to diverse climatic conditions. The study opens avenues for further exploration at the intersection of meteorology, spatial analysis, and machine learning. In-depth analysis into factors contributing to NN\_Model\_2's superior performance, such as its ability to handle non-linear relationships or incorporate additional precipitation influences, could be pursued. In this study, we focused exclusively on daily accumulated precipitation data and did not incorporate any sub-daily datasets. Furthermore, the temporal axis represents another dimension for interpolation or local projection. For datasets with higher temporal resolution, the temporal interpolation techniques can be utilized to achieve smoother transitions and more accurate representations of time-dependent variations.

Overall, this study establishes a critical first step towards democratizing weather information by enabling accurate local-scale approximation of meteorological variables. By bridging the gap between model forecasts data at grid points and location-specific needs, this technology empowers communities to proactively manage weather-related risks and policymakers to make informed decisions in real-time. This process empowers individuals to better understand their local weather patterns and ultimately mitigate the impact of unforeseen weather events on life and property. As we delve deeper into the interdisciplinary space of weather forecasting, we pave the way for enhanced understanding and prediction capabilities, fostering a future where weather information empowers, rather than hinders, individual and societal well-being.


\section*{Acknowledgements}
The IITM Pune is funded by Ministry of Earth Sciences, Government of India. The computational infrastructure for this study was provided by the HPC facility at IITM Pune. The authors gratefully acknowledge Dr. Udit Bhatia of IIT Gandhinagar for his insightful discussions.

\vspace{-2mm}
\section*{Funding information}\vspace{-4mm} No funding was received for conducting this study.
\vspace{-4mm}
\section*{ORCiD} \vspace{-4mm}BK: 0000-0001-7047-551X.  BBS: 0000-0003-3877-6800. RC: 0000-0002-1973-7307
\vspace{-4mm}
\section*{Conflict of interest}\vspace{-4mm}	The authors declare no competing interests.

\section*{References}


\begin{small}

\bibliographystyle{iopart-num}
\bibliography{nowcasting_NN_model_ref_n}
\end{small}


\section*{Supplementary material}\label{secA1}
\nopagebreak	
	\begin{figure*}[htpb]
		\centering
		\includegraphics[height=0.70\textheight, width=0.95\textwidth]{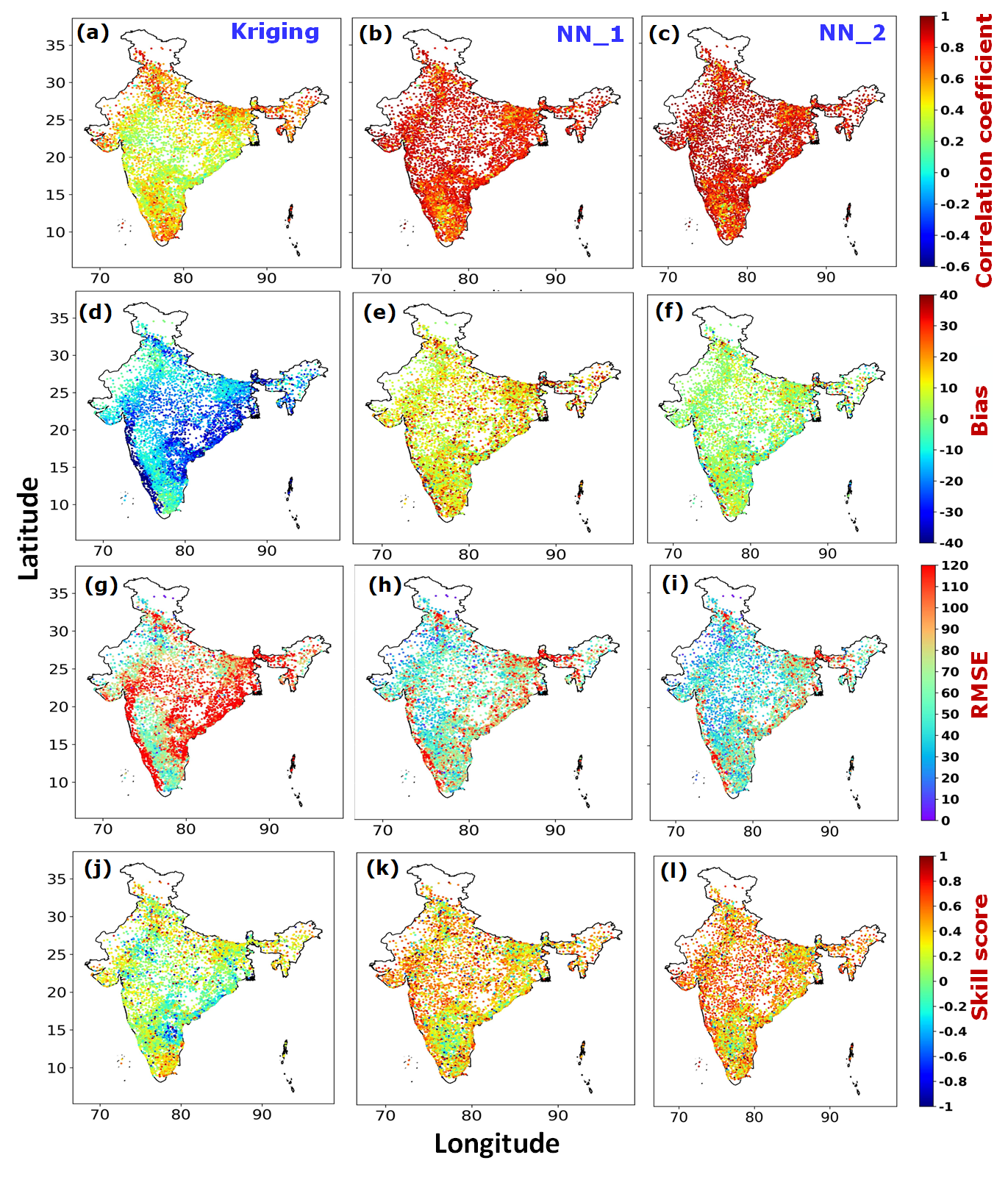}
		\caption{Spatial plots CC, biases, RMSE and skill score calculate from three models for whole year data. }
		\label{fig:spatial_plots_CC_RMSE_bias_whole_year}
	\end{figure*}
	

	\begin{figure*}[htpb]
		\centering
		\includegraphics[height=0.22\textheight, width=0.99\textwidth]{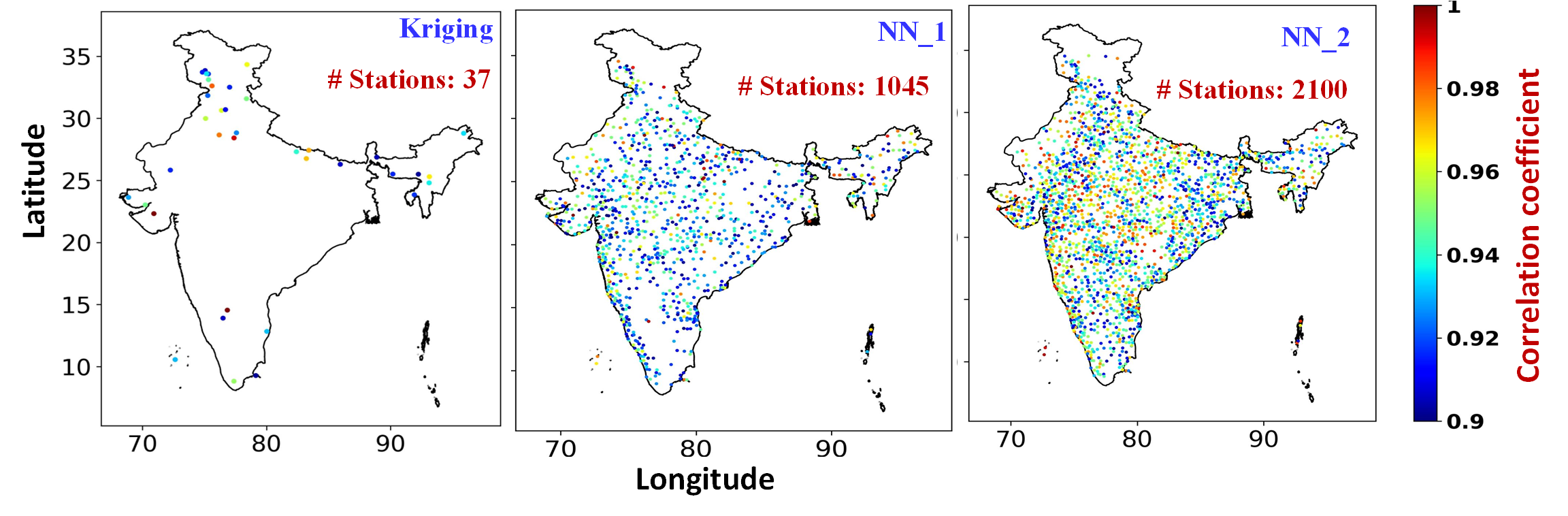}
		\caption{The station locations where the model performed  well using the whole year data, yielding very high correlation values. The total count of stations where the correlation coefficient (CC) value exceeded the threshold of 0.90 is provided for all three models are also indicated. }
		\label{fig:high_CC_whole_year}
	\end{figure*}



\end{document}